\documentstyle[12pt]{article}

\begin{document}

\title{Special Deformed Exponential Functions Leading to More Consistent
Klauder's Coherent States}  \author{}
\date{}
\maketitle

\center{\bf M. EL BAZ}\footnote{E-mail address: moreagl@yahoo.co.uk}

\center{\it Facult\'e des sciences, D\'epartement de Physique, LPT-ICAC, \linebreak Av. Ibn Battouta, B.P. 1014, Agdal, Rabat, Morocco}
\vspace{0.7cm}
\center{\bf Y. HASSOUNI}\footnote{E-mail address: Y-hassou@fsr.ac.ma}

\center{\it Facult\'e des sciences, D\'epartement de Physique, LPT-ICAC,
\linebreak Av. Ibn Battouta, B.P. 1014, Agdal, Rabat, Morocco \linebreak and
\linebreak the Abdus Salam International Centre for Theoretical Physics
\linebreak strada costiera 11 , 34100 Trieste, Italy}

\vspace{3cm}
\abstract{We give a general approach for the construction of deformed
oscillators. These ones could be seen as describing deformed bosons. Basing
on new definitions of certain quantum series, we demonstrate that they are
nothing but the ordinary exponential functions in the limit when the
deformation parameters goes to one. We also prove that these series converge
to a complex function, in a given convergence radius that we calculate. Klauder's
Coherent States are explicitly found through these functions that we design
by deformed exponential functions} 

\pagebreak

\section{Introduction}
The notion of coherent states (C.S.) saw its origins in the early times of 
quantum mechanics. In 1926 Schrodinger \cite{schro} introduced a set of wave
functions to describe some particular wave packets for the harmonic oscillators (H.O.).
After that, Van Neuman used these functions to investigate the coordinate and momentum measurment process in quantum theory. These ideas did not attract the attention of the eminent scientists for a long period. In the sixties, Glauber \cite{glauber}, who is considered as one of the fathers of the theory of C.S., designed these states by coherent states, he also
proved that they are adequate to describe a coherent laser beam in the
framework of quantum theories. Many works after that presented these states in a moremodern form \cite {klauder, per63, nieto}. In \cite{klabook} the authors defines the properties that are conserved in all these generalisations. This constitutes the minimum set of properties that a state should satisfy to be coherent. These properties are:
\begin{itemize}
\item Normalisability.
\item Continuity.
\item and Resolution of unity.
\end{itemize}
This last property is certainly the most important and the most restrictive
one as we shall see in this paper. In fact our aim here is to costruct C.S.
associated to deformed bosons, and the main difficulty preventing us from
presenting a general method of doing so, is exactly this property.

In the present work we discuss the problem of constructing C.S. associated to deformed bosons. We present a general approach of achieving this. In fact we deal with an algebra unifying all the known deformations of the H.O's algebra. We carry out the calculations for a particular case (of the unifying algebra) which is neverthless, itself a generalization of many deformations appeared  in the literature.

In the first part of this work are given some preleminaries, on the notions of C.S. and
deformed bosons, and some of their properties that will be useful in the
second part, where we discuss in more detail the construction of coherent
states for deformed bosons using a well defined sheme. The difficulty is
located indeed in finding the analogue of the exponential function ensuring
the obtention of the C.S. starting from the vacuum state in the Fock space. We
proceed by giving what we call in this section quantum series. These ones can
converge to a function that is viewed as a deformed exponential function in
our context. We calculate in detail the convergence radius, for which these quantum series could have a sense. This radius is compared with the one already
used in \cite{arik}. The last part of this paper is a sort of a recapitulation
where we give all the main ideas discussed here.

\section{Preliminaries}

\subsection{Coherent States}

According to Klauder \cite{klauder,klabook}, the minimum set of requirements to
be imposed on a state $|z>$, for to be a coherent state C.S. is:

\renewcommand{\theenumi}{\alph{enumi}} 
\begin{enumerate} 
\item {\it Continuity in the label} i.e.

\begin{equation}
\bigg | |z> -|z'>\bigg |^2 \;\longrightarrow 0 \;\;\;\;\; when 
\;\;\;\;\;|z-z'|^2 \longrightarrow 0 
\end{equation} 

\item {\it Resolution of unity:}

The states $|z>$ must provide a decomposition (not necessarily unique) of
the identity operator:
\begin{equation}
\int \int d \mu (z) |z><z| = I
\end{equation}
where $d\mu (z)$ is a measure in the label space.
\item {\it Normalisability}
\begin{equation}
<z|z> = 1
\end{equation} 
this last condition can be imposed, almost always, without affecting the first
two requirements. However when this is not the case this condition should be
droped \cite{klabook}
\end{enumerate} 

In general, the first and third conditions are easily satisfied. This is not
the case for the second. In fact, this condition restricts considerably the
choice of the states to be considered.

The most familiar case is, of course, the case for which the C.S. were first
constructed i.e. the "bosonic" Harmonic Oscillator (H.O.) and the so-called
canonical C.S. associated to it:

\noindent We recall that the bosonic H.O.'s annihilation and creation
operators satisfy the commutation relation $[a,a^+] = 1$, and the canonical
C.S. are given by:

\begin{eqnarray}
|z> &=& \exp(-{|z|^2 \over 2}) \exp(za^+) |0> \nonumber \\
 &=& \exp(-{|z|^2 \over 2}) \sum _{n \ge 0} {z^n \over {\sqrt{n!}}}|n>
\end{eqnarray}      
where $|n>$; $n=1,2....n$ are the usual number states generated from the
vacuum state $|0> $ ($a|0> = 0$) using:
\begin{equation}
|n> = { {(a^+)}^n \over \sqrt{n!}}|0>
\end{equation}
These states do satisfy all the requirements mentionned above, in particular
they provide a resolution of unity (2) with 
\begin{equation}
d\mu (z) = d^2z W(|z|^2)
\end{equation}
where $d^2z = dRe(z)\;dIm(z)$ and the weight function is $ W(|z|^2) = {1
\over \pi} $

\subsection{Deformed bosons}

One of the most trivial ways of generalising the concept of C.S. is to
construct such states for deformed bosons \cite{arik}. In fact, during the last
decades many deformations of the usual H.O. appeared in the litterature
\cite{greenberg, macfarlane}. In a previous work \cite{1}, we have discussed
the possibility of unifying all these deformations. We have also presented an
algebra ${\cal A}_Q$ that seems to englobe most of this cases.

The algebra ${\cal A}_Q$ is generated by the triplet $\{a, a^+, I \}$, and is
defined through the following "Q-mutation" relations:

\[ 
[a,a^+]_Q  =  aa^+ - Qa^+a = \Delta'_Q  
\] 
\begin{equation} 
[a,\Delta _Q]_Q  =  a\Delta _Q - Q\Delta _Qa = \Delta'_Qa  
\end{equation} 
\[ 
[a^+,\Delta _Q]_Q =  a^+\Delta_Q - Q\Delta _Q a^+ = -a^+\Delta' _Q
\] 
\hspace{6cm}$ \vdots  $ 

\noindent where $Q$ is a complex parameter (of deformation), $\Delta _Q =
a^+a$, and $ \Delta _Q'$ is to be interpreted as a "Q-derivative" of $\Delta
_Q$.

The Fock space basis associated is defined in the usual way: Given a vacuum
state $|0>$; ($a|0> = 0$) the different number states are generated through the
action of the creation operator on this state:
\begin{equation}
(a^+)^n |0> = \sqrt{[n]_Q!} |n>
\end{equation}
where the factoriel function is defined as:
\[
[n]_Q! = [n]_Q [n-1]_Q....[1]_Q
\]
\[
[0]_Q!=1
\]
and the function $[n]_Q$ appears in:
\begin{eqnarray} 
a|n> & = & \sqrt{[n]_Q} \;|n-1> \nonumber \\ 
a^+ |n> & = & \sqrt{[n+1]_Q} \; |n+1> \nonumber \\ 
\Delta_Q |n> & = & [n]_Q \; |n> \\ 
\Delta'_Q |n> & = & ([n+1]_Q -Q[n]_Q) \; |n> \nonumber 
\end{eqnarray} 

This algebra could be seen as generalising all the particular deformed boson's
algebras. These can indeed be obtained from it by choosing the adequate function
$[n]_Q$. In particular in \cite{1} we have discussed the case which is obtained from the algebra ${\cal A}_Q$ by choosing:
\begin{equation}
[n]_Q = { Q^n - Q^{-n} \over Q-Q^{-1}}
\end{equation}
which leads to the following "Q-mutation" relation:
\begin{equation}
aa^+ - Qa^+a = Q^{-N}
\end{equation}
where $N$ is the number operator defined by:
\[
N|n> = n|n>
\]
\section{C.S. for Deformed Bosons}

Let's now turn to the problem of constructing C.S. associated to these
deformed H.O. This problem was treated for most
of the deformations existing in the litterature. However, presenting a
general method for constructing C.S. associated to the algebra ${\cal A}_Q$ in
its general form is not possible as was demonstrated in \cite{1}.

The main difficulty in doing so resides; in a large proportion; in the
fulfillement of condition (b) i.e. in finding the adequate resolution of unity.
Concerning this point we distinguish two main approachs in the litterature:

\begin{itemize}
\item Deforming the concept of integration and differentiation in such a manner
that these satisfy similar properties of those obeyed by ordinary ones. This
was done in \cite{arik, per96} for the particular case where $ [n] = {q^n - 1 \over q - 1}$, this corresponds to the conventional quons $ aa^+ - qa^+a =1$. Achieving this, for the algebra ${\cal A}_Q$ in its general form is still missing.

\item Leaving intact the concept of integration and differentiation, and try
to fulfill (b) in an analogous manner as in (6) i.e. find the appropriate
weight function, such that (b) holds \cite{pen1, pen2, pen3}
\end{itemize}

Using this last method we have succeeded in \cite{1} to construct C.S. for
(10). However the construction was valid only when the parameter of the
deformation is a root of unity. We now investigate on the possibilities of
extending this construction.

As in \cite{1} we begin by constructing C.S. candidates for the algebra ${\cal
A}_Q$ in its general form, then specify the particular algebra we are
interested in. Such states are given by:
\begin{equation}
||Q,z> = \sum _{n\ge 0} { z^n \over \sqrt{[n]_Q!}}\; |n>
\end{equation} 
These states are, of course, eigenstates of the annihilation operator
\[
a ||Q,z> = z \, ||Q,z>
\]

We introduce the deformed exponential function associated to the algebra
${\cal A}_Q$ by noticing that (12) can be rewritten as:
\begin{eqnarray}
||Q,z> &=& \sum _{n\ge 0}{z^n \over [n]_Q!}\; (a^+)^n |0> \nonumber \\
&\equiv & \exp ^{(1)}_Q (za^+) |0>
\end{eqnarray}
The use of the superscript (1) will become clear in what follows.

Let's now see in which circumstances these vectors are C.S. in
the sens of Klauder.

First of all these states are clearly normalisable:

\begin{eqnarray}
<Q,z||Q,z> &=& \sum _{n \ge 0} {|z|^{2n} \over \sqrt
{ \bar{[n]}_Q![n]_Q!}}\nonumber \\  
&=& \sum_{n \ge 0} { |z|^{2n} \over | [n]_Q |!} \\ 
&\equiv & \exp ^{(2)}_Q (|z|^2) \nonumber
\end{eqnarray}
where the bar means complex conjugation, and we have introduced another
deformed exponential function as follows\footnote{ In \cite{1} we have imposed
on $[n]_Q$ to satisfy $\bar{[n]}_Q = [n]_Q$, which means then $ \exp ^{(1)} _Q =
\exp ^{(2)} _Q $ }: 
\begin{equation} 
\exp ^{(2)}_Q (x) = \sum _{n\ge 0} { x^n \over |[n]_Q|! } 
\end{equation}

We point out at this step, that in order to continue, the "quantum series"
defining the two exponential functions introduced, must converge, and have a
non nul radius of convergence. For the moment we impose this condition, we
shall return to this point later.

From here and after we shall deal only with the normalized states:
\begin{eqnarray}
|Q,z> &=& {\cal N}(|z|^2) ||Q,z> \nonumber \\
&=& \bigg [ \exp ^{(2)} _Q ({|z|^2})\bigg ]^{-{1 \over 2}} ||Q,z> \\
&=& \bigg [ \exp ^{(2)} _Q ({|z|^2})\bigg ]^{-{1 \over 2}} \exp ^{(1)} _Q (za^+)|0> \nonumber
\end{eqnarray}

The overlap term of two of these states is given by:

\begin{equation}
<Q,z|Q,z'> = {\cal N} (|z|^2) {\cal N} (|z'|^2) \; \exp ^{(2)} _Q (\bar z z') 
\end{equation}
This equation, together with:
\begin{equation}
\bigg | |Q,z> - |Q,z'>\bigg |^2 = 2(1-Re <Q,z|Q,z'>)
\end{equation}
implies the continuity of the states (16) in their label $z$.

So far, we have seen that conditions (a) and (c) haven't, almost, imposed any
restrictions on the choosen states. In what follows we will see that this is
not the case with condition (b).

The question is, in which cases the states (16) do provide a resolution of
unity (2)? We shall try to respond to this question in the case (6).

Using (16), (12) and (6) in (2) we obtain:
\begin{equation} 
\int \int _{|z|^2 < R_Q} d^2z \sum _{n,m\ge 0}{z^n\over \sqrt {[n]_Q!}}
{{\bar z}^m \over \sqrt{\bar{[m]}!}}{\cal N}^2({|z|}^2)|n><m| W({|z|}^2) =
I  
\end{equation}

where $ d^2z=d\alpha d\beta = rdrd\theta $ when  $z=\alpha + i\beta = r
e^{i\theta}$ and $ R_Q$ is the convergence radius of the series in (15).

\noindent from which we obtain:
\begin{equation} 
\sum _{n\ge 0} { \pi \over |[n]_Q|!} \bigg \{ \int _{0}^{R_Q} dx\; x^n \; {\cal
N}^2(x) W(x) \bigg \} |n><n| = I  
\end{equation}
where $ x=r^2$.

The completness of the states $|n>$ implies:
\begin{equation}
\int _0 ^{R_Q} dx\; x^n \; {\cal N}^2(x) W(x) = {|[n]_Q|! \over \pi }
\end{equation}

putting
\[
\tilde W(x) = {\cal N}^2(x) W(x)
\]

we get
\begin{equation} 
\int dx \;x^n \; \tilde{W}(x) = {{|[n]_Q}|! \over \pi } 
\end{equation} 
which is the well known power moment problem when $R_Q = \infty $, or the
Stieljes moment problem when $ R_Q < \infty $ \cite{kps, akh}

It is clear that a general solution for this
equation (i.e. for a generic r.h.s) can not be given \cite{kps, akh}.

We propose to solve this equation using the Fourrier transforms method:

multiplying equation(22) by $\bigg ({(iy)^n \over n!} \bigg )$ and summing
over $n$ yields:

\begin{eqnarray}
\int _0 ^{R_Q} dx \; e^{iyx} \tilde W(x) &=& \sum _{n \ge 0} { |[n]_Q|! (iy)^n
\over \pi n!} \\
&=& \bar W(y) \nonumber
\end{eqnarray}

To proceed, the series in the r.h.s should converge. This imposes a severe
restriction on the $[n]_Q$'s to be choosen, which means on the corresponding
particular algebra ${\cal A}_Q$.

In \cite{1} we have been interested in the case (10), we present here the
following proposition concerning this choice:
\vspace{0.4cm}

{\bf Proposition1:}

The series in (23) with $[n]_Q = {Q^n - Q^{-n} \over Q - Q{-1}}$ converges
only when $|Q| = 1$. It diverges otherwise.

\vspace{0,4cm}

Too restricted as a choice one could say!

To overcome this situation we introduce another parameter of deformation $p$. This is equivalent to see the parameter $Q$ as composed of two parameters $q$
and $p$.

We define the function $[n]_{q,p}$ as follows:
\begin{equation}
[n]_{q,p} = { q^n - p^{-n} \over q - p^{-1}}
\end{equation}

This is clearly a generalisation of the first case (10), but it stills a
particular case of the algebra ${\cal A}_Q$

The operators $a$ and $a^+$ in this case satisfy:

\begin{equation}
aa^+ - qa^+a = p^{-N}
\end{equation}

To construct the C.S. associated to this, we have to change the label
$Q$ appearing in all previous steps by $q,p$! For convenience we rewrite the
main functions here.

\noindent The two deformed exponential functions:
\begin{eqnarray}
\exp ^{(1)} _{q,p} (x) &=& \sum _{n\ge 0} {x^n \over [n]_{q,p}!} \\
\exp ^{(2)} _{q,p} (x) &=& \sum _{n\ge 0} {x^n \over |[n]_{q,p}|!}
\end{eqnarray}
and the series $ \bar W(y)$ in (26) will become
\begin{equation}
\bar W(y) = \sum _{n\ge 0} {|[n]_{q,p}|! (iy)^n \over \pi n!}
\end{equation}
and we have the following result
\vspace{0,4cm}

{\bf Proposition2:}

The series in (26), (27) and (28) converges simultaneously in two cases:

(i) $ \;|q|\le 1$ and $|p| = 1$

(ii) $|q|=1$ and $|p| \ge 1$

otherwise, at least one of these series diverges.

\vspace{0,4cm}

This proposition makes it clear: to continue we shall take our parameters as
in $(i)$ or $(ii)$. It is also clear that when this is the case, the two
deformed exponential functions (26) and (27) converges for\footnote {for
$q=p$ we obtain the case considered in\cite{1}, i.e. (11), and in that case $
R_q = \infty$.} $|x| < |q - p^{-1}|^{-1} = R_{q,p}$

Now we return back to equation (23), for the case we are considering, the
series $\bar W(y)$ converges. Thus the inverse Fourrier transform of $\bar
W(y)$ exists and is given by:
\begin{equation} 
\tilde W(x) = { 1 \over 2\pi} \int _{-\infty}^{\infty}{e^{-iyx} \bar W(y) dy} 
\end{equation}
where $x<R_{q,p}$ is overtone.

\noindent and we get the weight function:
\begin{equation}
W(x) = {{\cal N} ^{-2}(x) \over 2\pi} \int _{-\infty}^{\infty}{e^{-iyx} \bar
W(y) dy} 
\end{equation}
A resolution of unity, thus exists for the case we considered (24) ( with the
conditions on $q,p$ stated on Proposition2 ), which in turn means that indeed
the vectors (16) are coherent states for this system.

\section{ A Recapitulatory}

In this paper we have discussed the construction of C.S. associated to
deformed bosons oscillators. Since all these deformations can be extracted
from the algebra $ {\cal A}_Q$ (7),(9), it will be more intersting to achieve
such a construction for this algebra. We have tried to do this, using one of
the most trivial ways for constructing C.S. i.e. (12) and (6). We failed in
reaching our goal due to two main problems. The first is due to the
restrictions that imposes the resolution of unity (2) and the equations that
derives from it (22) and (23). The other problem is related to the
deformed exponential functions introduced. In fact the two series defining
this functions (13) and (15) must converge and have a non nul convergence
radius.

The series in (23) does not converge for all the $[n]_Q$'s that one takes, so is
the case for (13) and (15), it is thus impossible to construct C.S. for
${\cal A}_Q$ in its general form using the approach we've taken in this
paper. However using this same approach we succeded to do so for a particular
case, namely when $[n]_Q = [n]_{q,p} = { q^n - p^{-n} \over q - p^{-1} }$.
It's true that this still a particular cases of ${\cal A}_Q$, but this
deformation generalises at least 3 particular cases:

\begin{itemize}
\item when $p \longrightarrow q$ we obtain (10) and (11).

\item when $p \longrightarrow 1$ we obtain the conventional
quons \cite{greenberg}.

\item when $p \longrightarrow 1$ and $q \longrightarrow 1$ we obtain the
ordinary bosons.
\end{itemize}

It will be interesting to investigate on the possibility of achieving our goal
(i.e. C.S. for ${\cal A}_Q$) using other approachs, for instance changing our
starting point (12), or use the other method mentioned above (begining of
section 3) concerning the fulfillement of (2).

\section*{ Aknowledgment }
The authors wishes to thank the International Center for Theoretical Physics, I.C.T.P. where part of the work was achieved.

\end{document}